# STUDY OF THE UNILAMELLAR VESICLE STRUCTURE VIA SANS ON THE BASE OF THE SFF MODEL

E.V. Zemlyanaya[1], M.A. Kiselev[2,3], J. Zbytovska[3], L. Almasy[4], T. Gutberlet[5], P. Strunz[5], S. Wartewig[6], G. Klose[7], R.H.H. Neubert[4]

[1] - Laboratory of Information Technology, Joint Institute for Nuclear Research, Dubna, Russia
[2] - Frank Laboratory of Neutron Physics, Joint Institute for Nuclear Research, Dubna, Russia
[3] - Institute of Pharmaceutical Technology and Biopharmacy, Martin-Luther-University, Halle (Saale), Germany
[4] - Budapest Neutron Scattering Center, Hungary
[5] - Paul Scherrer Institute, Villigen, Switzerland
[6] - Institute of Applied Dermatopharmacy, Martin-Luther-University, Halle (Saale), Germany
[7] - Physical Faculty, Leipzig University, Germany

**Introduction**

Investigation of model stratum corneum (MSC) lipid structure is a subject of interest in pharmacology and dermatopharmacy as a background for the formulation of new enhancer of drug penetration through human stratum corneum [1].

The determination of the internal lipid membrane structure is far beyond the resolution of light microscopy or light scattering [1]. Neutron scattering in the range of wavelength 1-10 Å provides better spatial resolution of an internal structure on the scale of Ångströms. It allows determine the internal membrane structure with reasonable accuracy. Membrane thickness, thickness of polar head groups, thickness of methyl and methylene groups, surface area per lipid molecules, thickness of hydrophilic and hydrophobic membrane regions can be calculated from scattering experiment. The spatial resolution of neutron scattering experiments mainly depends on the possibility to collect a scattering curve in the wide range of scattering vector [2-4]. The aim of the experiment is to determine the form factor of the membrane with maximum accuracy and further apply the model methods to calculate a membrane structure [5].

Three experimental methods can be used for this purpose, dependently on the possibility to prepare the relevant samples: neutron diffraction on the oriented multilamellar samples on the solid substrate, neutron small-angle scattering (SANS) of highly diluted vesicular systems (1-10% w/w), and neutron reflectivity on the lipid bilayer on the silicon substrate at water excess.

Information about the internal membrane structure can be obtained from neutron diffraction experiments on oriented planar multilamellar systems, which are partially dehydrated [7]. This method is more appropriate for the modulation of physiological conditions of human skin. The experimental resolution depends on the number of detected diffraction peaks, connected with possibility to prepare the highly oriented multilamellar system on the solid substrate.

Small-angle neutron scattering is the best method to study a membrane structure of the highly diluted unilamellar or oligolamellar vesicular systems. In the case of MSC study, the small-angle scattering technique can be applied for systems where the possibility to prepare highly oriented planar multilamellar systems is limited, or as the complementary method. First efforts to investigate the internal membrane structure of oligolamellar vesicles via SANS gave about 1.5 Å accuracy of the membrane thickness determination, accuracy of 2 Å of the hydrophobic core determination, and new information about membrane hydration [4].

The separated form factors approach (SFF) was developed for the characterization of the unilamelar vesicles with diameter ≥ 500 Å via SANS [8-11]. The application of fluctuated membrane model in the framework of the SFF shows that scattering density from $D_2O$ distribution function inside of hydrophilic region dominates under that from polar head groups [12]. Hence, one can expect that the hydrophilic-hydrophobic model (HH-model) with the linear $\rho(x)$ function of the neutron scattering length density across membrane in the hydrophilic region can provide realistic and simple approximation of the lipid membrane structure. In the HH-model, only two fit parameters (thickness of membrane d, and thickness of hydrophobic region D) are needed to simulate ρ(x) function. This approach sufficiently improves the fitting convergence relative to the more complex models of ρ(x). The same approximation of the scattering length neutron density ρ(x) has been applied in [6] for an oligolamellar POPC vesicles study [13].

In present study, the parameters of the polydispersed unilamellar vesicle population are analyzed on the basis of the separated form-factors model for three different type of membranes. The neutron scattering length density across the membrane is simulated on the base of the HH-model with the linear water distribution.

**Experiment, sample preparation, and methods**

SANS measurements were carried out at YuMO time-of-flight spectrometer in JINR (Dubna), at Yellow Submarine spectrometer at study state reactor in Budapest, and at SANS-I spectrometer of spallation neutron source at PSI (Switzerland).

Unilamellar vesicles were prepared via extrusion of multilamellar vesicle suspension (liposomes) in $D_2O$ through polycarbonate filters with pore diameter 500 Å.

SFF model was applied to describe a vesicle form and internal membrane structure. This approach allows to simulate the scattering length density of neutrons across the bilayer by any integrable function [8]. In this study, the scattering length density ρ(x) across the bilayer was simulated by the HH approximation (see Fig. 1). The scattering density of hydrocarbon chains $\rho_{CH}$ was fixed as $-0.36 \cdot 10^{10}$ см$^{-2}$, $\rho_{D2O} = 6.4 \cdot 10^{10}$ см$^{-2}$ [4]. Parameters of the DMPC vesicle population (average radius <R>, polydispersity σ, membrane thickness d, thickness of hydrophobic part D) were restored only from the SANS spectra, without additional experimental methods (light scattering, diffraction, etc.). Additionally, in the case of known molecular volume, the number of water molecules per one lipid molecule in the membrane bilayer $N_w$ and area per one lipid molecule A can be calculated.

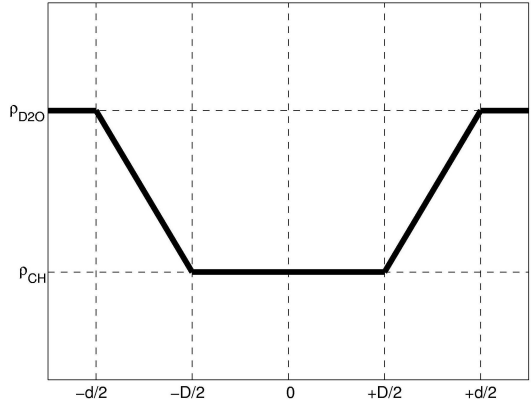

*Fig. 1. HH model of scattering length density across the lipid bilayer. d is a membrane thickness, D is a thickness of the hydrophilic part of the membrane*

Accordingly the SFF model, the macroscopic cross section of the monodispersed vesicle population is equal

$$\frac{d\Sigma}{d\Omega_{mon}}(q) = n \cdot F_s(q,R) \cdot F_b(q,d) \cdot S(q) \quad (1)$$

where S(q) is the vesicle structure factor in Debye form [10], n – number of vesicles per unit volume, form factors of sphere and bilayer are, respectively:

$$F_s(q,R) = \left(4\pi \cdot \frac{R^2}{qR} \cdot \sin(qR)\right)^2 \quad (2)$$

$$F_b(q,d) = \left( \int_{-d/2}^{d/2} \rho(x) \cdot \cos(qx) \cdot dx \right)^2 \quad (4)$$

The vesicle polydispersity is described by nonsymmetric Schulz distribution

$$G(R) = \frac{R^m}{m!} \cdot \left( \frac{m+1}{<R>} \right)^{m+1} \cdot \exp\left[ -\frac{(m+1) \cdot R}{<R>} \right] \quad (5)$$

Relative standard deviation of vesicle radius is

$$\sigma = \sqrt{\frac{1}{(m+1)}} \quad (6)$$

Macroscopic cross section $d\Sigma(q)/d\Omega$ of polydispersed vesicle population

$$\frac{d\Sigma}{d\Omega}(q) = \frac{\int_{R\min}^{R\max} \frac{d\Sigma}{d\Omega_{mon}}(q,R) \cdot G(R,<R>) \cdot dR}{\int_{R\min}^{R\max} G(R,<R>) \cdot dR} \quad (7)$$

should be corrected to the to the instrument resolution function

$$I(q) = \frac{d\Sigma}{d\Omega}(q) + \frac{1}{2} \cdot \Delta^2 \cdot \frac{d^2 I_m(q)}{dq^2} \quad (8)$$

where $\Delta^2$ is a second moment of the resolution function [14].

The fitting parameters are average vesicle radius $<R>$, coefficient of polydispersity m, number of vesicles n, and parameters of function $\rho_C(x)$, simulating the neutron scattering length density. One can also consider the incoherent background IB as another unknown parameter.

The fit quality was estimated as:

$$R_I = \frac{1}{N} \cdot \sum_{i=1}^{N} \left( \frac{|\frac{d\Sigma}{d\Omega}(q_i)| - |\frac{d\Sigma}{d\Omega}_{exp}(q_i)|}{|\frac{d\Sigma}{d\Omega}_{exp}(q_i)|} \right)^2 \times 100\%$$

**Results and discussions**

The scattering curve from polydispersed DMPC vesicle population was collected at YuMO spectrometer in the q range from $q_{min}=0.0084$ Å$^{-1}$ to $q_{max}=0.2$ Å$^{-1}$. Incoherent background was fixed parameter IB=0.00546 cm$^{-1}$. The scattering curve from the same sample was collected at SANS-I spectrometer at PSI in the q range from $q_{min}= 0.0033$ Å$^{-1}$ to $q_{max}=0.56$ Å$^{-1}$. Incoherent background was a free parameter in this case. Calculated vesicles parameters are presented in Table 1. Parameters obtained from the both instruments are in good agreement, but an accuracy of the parameter evaluation is better from PSI-I spectra because of wide q-range at SANS-I instrument. Important advantage of SANS-I spectrometer is the possibility to collect data at q>0.2Å$^{-1}$ and fit the value of incoherent background IB.

The membrane parameters derived from the unilamellar vesicles: d=49.6Å, D=18.6Å, A=61.4Å2, and $N_w$=14.1 are sufficiently different from that obtained from X-ray diffraction on the multilamellar vesicles: d=44.2Å, D=26.2Å, A=59.6Å$^2$, and $N_w$=7.2, except of the A value [15]. The increase of membrane curvature increases the bilayer hydration and membrane thickness [13]. The area per one DMPC molecules A looks as constant parameter for unilamellar and multilamellar vesicles, that reflects the permanent intermolecular interaction in the lateral directions. Probably, membrane hydration decreases during the aggregation of the unilamellar vesicles to the multilamellar and $N_w$=7.2 corresponds to the hydrated multilamellar vesicular system that is in the thermodynamic equilibrium with bulk water.

*Table 1. Parameters of 1% (w/w) DMPC unilamellar vesicle population at T=30$^o$C.*

|  | <R>, Å | σ,% | d, Å | D, Å | $N_w$ | A, Å$^2$ | IB, cm$^{-1}$ | $\chi^2$ | $R_I$ |
|---|---|---|---|---|---|---|---|---|---|
| YuMO Dubna | 277±5 | 31 | 49±2 | 20±3 | 12±3 | 60±4 | 0.00546 | 1.1 | 5·10$^{-5}$ |
| SANS I PSI | 275.6 ±0.5 | 27 | 49.6 ±0.2 | 18.6 ±0.4 | 14.1 ±0.3 | 61.4 ±0.4 | 0.00656 ±0.00003 | 6.6 | 5.9·10$^{-6}$ |

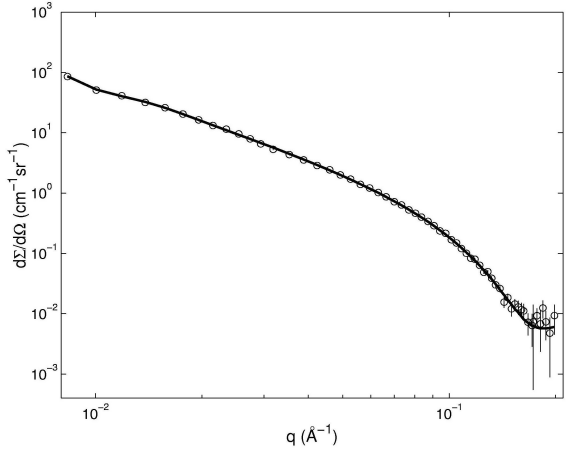 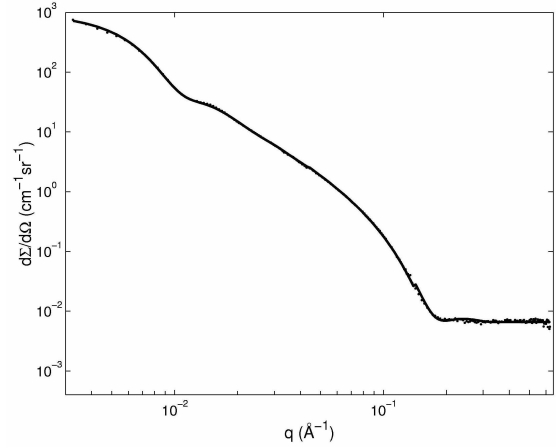

*Fig. 2. YuMO SANS spectrometer. Experimental and fitting curves for DMPC vesicles at T=30$^o$C. 1% DMPC concentration in $D_2O$.*

*Fig. 3. SANS-I PSI spectrometer. Experimental and fitting curves for DMPC vesicles at T=30$^o$C. 1% DMPC concentration in $D_2O$.*

The developed methods of the membrane structure analyses for one component membrane were applied to study the mixed binary and quaternary systems. The SANS experiment for the investigation of MSC membrane structure was carried out at the Budapest Neutron Scattering Center. The SANS spectrum obtained for vesicles composed of binary DMPC/Ceramide 3 and quaternary Ceramide 6/Cholesterol/Palmitic acid/Cholesterol Sulfate are presented at Fig. 4 and Fig. 5, respectively. The composition of binary DMPC/Ceramide 3 mixture corresponds to the ratio 6/1 in mole. Component ratio of Ceramide 6/Cholesterol/Palmitic acid/Cholesterol Sulfate is 55/20/15/10 (w/w).

The incoherent background was fixed IB=0.0109 cm$^{-1}$ for the case of binary DMPC/ceramide 3 vesicles; and IB was a free parameter for the case of quaternary Ceramide 6/Cholesterol/Palmitic acid/Cholesterol Sulfate vesicles. Obtained results are summarized in Table 2.

*Table 2. Parameters of mixed MSC vesicles at T=30$^o$C.*

|  | <R>, Å | σ, % | d, Å | D, Å | IB, cm$^{-1}$ | $\chi^2$ | $R_I$ |
|---|---|---|---|---|---|---|---|
| Binary MSC system | 256±5 | 22 | 50±3 | 27±5 | 0.0109 | 2.2 | 2·10$^{-4}$ |
| Quaternary MSC system | 349±7 | 39 | 48±10 | 29±4 | 0.0417±0.0002 | 1.2 | 1.2·10$^{-5}$ |

Ceramide 3 molecules increase (on the 8.4 Å) the thickness of the hydrophobic membrane region to that of pure DMPC. Hydrophobic region thickness of Cer3/DMPC membrane (27±5Å) is close to that of quaternary system Cer6/Chol/PA/CS (30±4Å) Probably, ceramide molecules mainly determine the thickness of the hydrophobic region.

Membrane thickness d=48Å and hydrophobic thickness D=29Å obtained from SANS for quaternary membrane Ceramide 6/Cholesterol/Palmitic acid/Cholesterol Sulfate with component ratio 55/20/15/10 (w/w) are in good agreement with values d=45.6Å and D=28Å evaluated from the neutron diffraction experiment on the same system with component 55/25/15/5 [16].

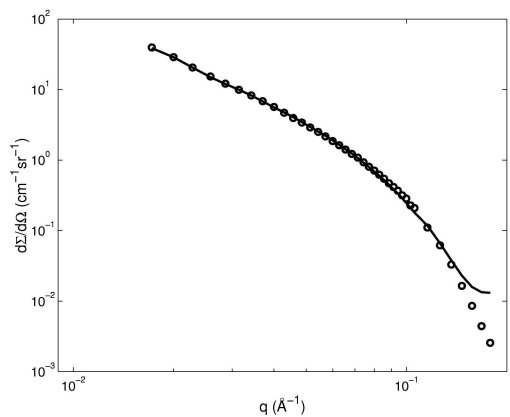
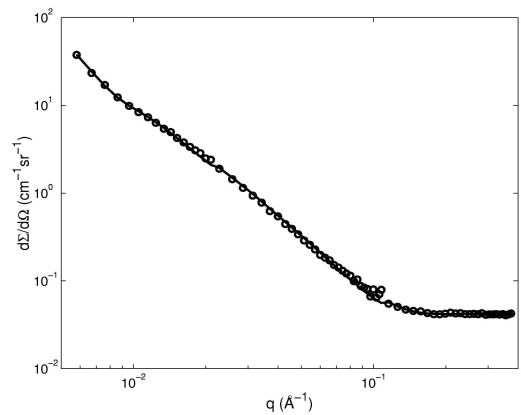

*Fig. 4. "Yellow Submarine" SANS spectrometer. Experimental and fitting curves for DMPC/ceramide 3 vesicles at $T=30^oC$. 2% (w/w) of common lipids concentration in $D_2O$.*

*Fig. 5. "Yellow Submarine" SANS spectrometer. Experimental and fitting curves for Ceramide 6/Cholesterol/Palmitic acid/Cholesterol Sulfate vesicles at $T=30^oC$. 0.5% of common lipid concentration in $D_2O$.*

**Acknowledgements**

This work was supported by a grant of the Federal State of Saxony-Anhalt (project 3482A/1102L) and the Ministry of Industry, Science and Technology RF (contract № 40.012.1.1.1148), and RFBR (grant 03-01-00657).